\documentclass[conference]{IEEEtran}
\IEEEoverridecommandlockouts

\usepackage{cite}
\usepackage{amsmath,amssymb,amsfonts}
\usepackage{algorithmic}
\usepackage{graphicx}
\usepackage{textcomp}
\usepackage{xcolor}
\def\BibTeX{{\rm B\kern-.05em{\sc i\kern-.025em b}\kern-.08em
    T\kern-.1667em\lower.7ex\hbox{E}\kern-.125emX}}

\usepackage{caption}
\usepackage{subcaption}
\usepackage[draft]{hyperref}       
\usepackage{url}            
\usepackage{booktabs}       
\usepackage{amsfonts}       
\usepackage{nicefrac}       
\usepackage{tikz}
\usepackage{tabularx,booktabs}
\usepackage{amsmath,amssymb,amsfonts}
\usepackage{enumitem}
\usepackage{algorithm}
\usepackage{algorithmic}
\newtheorem{definition}{Definition}

\newcommand{\flow}{flow}

\newcommand{\jump}{jump}
\newcommand{\pred}{pred}

\begin{document}

\title{A Passive Online Technique for Learning \\ Hybrid Automata from Input/Output Traces
}
\author{\IEEEauthorblockN{Iman Saberi}
\IEEEauthorblockA{\textit{Department of Electrical and Computer Engineering} \\
\textit{University of Tehran}\\
Tehran, Iran \\
iman.saberi@ut.ac.ir}
\and
\IEEEauthorblockN{Fathiyeh Faghih}
\IEEEauthorblockA{\textit{Department of Electrical and Computer Engineering} \\
\textit{University of Tehran}\\
Tehran, Iran \\
f.faghih@ut.ac.ir}

\and

\IEEEauthorblockN{\centerline{Farzad Sobhi Bavil}}
\IEEEauthorblockA{\textit{Department of Electronic Research and Innovation} \\
	\textit{Crouse Company}\\
	Tehran, Iran \\
	f.sobhi@crouse.ir}}

\maketitle

\begin{abstract}
	
Specification synthesis is the process of deriving a model from the input-output traces of a system. It is used extensively in test design, reverse engineering, and system identification. One type of the resulting artifact of this process for cyber-physical systems is hybrid automata.  They are intuitive, precise, tool independent, and at a high level of abstraction, and can model systems with both discrete and continuous variables. In this paper, we propose a new technique for synthesizing hybrid automaton from the input-output traces of a non-linear cyber-physical system. Similarity detection in non-linear behaviors is the main challenge for extracting such models. We address this problem by utilizing the Dynamic Time Warping technique. Our approach is passive, meaning that it does not need interaction with the system during automata synthesis from the logged traces; and online, which means that each input/output trace is used only once in the procedure. In other words, each new trace can be used to improve the already synthesized automaton. We  evaluated our algorithm in two industrial and simulated case studies. The accuracy of the derived automata show promising results.

\end{abstract}


\section{Introduction}
Learning the behavior of  systems has been  growing significantly over the last few years.  The behavior of a system could be represented by different artifacts, such as automata~\cite{medhat2015framework,soto2019membership} or formal languages~\cite{jin2015mining,kong2014temporal,jha2019telex}. Automata-based models are precise and intuitive artifacts. These tool-neutral models provide an abstract comprehension of the system's behavior. Traditionally, automata-based models are constructed by software engineers as a system specification artifact. However, in some cases, the engineers have a system without a model, and need to derive an automaton representing its behavior.  With the advances emerging in statistical analysis and process mining areas, there are  approaches proposed in this field that automatically mine these artifacts by observing and analyzing the input-output traces of the System Under Learning (SUL).

Traditional automata-based models, such as DFAs or NFAs are versatile artifacts for modeling systems with  discrete state space. They have been extensively used for modeling in areas, such as web services~\cite{raffelt2008hybrid} or network protocols~\cite{aarts2015generating,fiteruau2016combining}. However, today's software systems mostly consist of both continuous and discrete state space. Typically, these systems evolve continuously during the time, until an event happens, which takes them to another state, where the behavior of their continuous part changes differently. The behavior of this type of systems could be modeled by {\em hybrid automata}. 

Learning these models has several advantages such as understanding the behavior of complex systems~\cite{weiss2018extracting,okudono2020weighted}, automatically generation of software specifications and source codes~\cite{esparza2011learning,aarts2015generating}, model-based testing of systems without models~\cite{choi2013automated,choi2013guided}, and verification and validation~\cite{fiteruau2016combining}. Moreover, since these models predict the system's outputs in an explainable and transparent manner, learning them could be considered as an explainable machine learning method.
 For example, the authors in~\cite{weiss2018extracting} extract DFA from recurrent neural networks (RNNs) and provide an interpretable model of RNNs.   In~\cite{lamrani2018hymn}, the authors exploit Fisher Information analysis and Cramer Rao bound theorem to construct a linear hybrid automaton and model the behavior of an Artificial Pancreas.

\noindent We can categorize approaches for auotmata learning in two different ways~\cite{maier2014online}:
	\begin{enumerate}
		\item \textbf{Active vs passive}:
		Active learning algorithms directly interact with  SUL, and hence, they could request any traces of inputs, along with their outputs.  In passive learning, on the other hand, algorithms can only employ available traces of a system to synthesize an automaton.

		\item \textbf{Online vs offline}:
		Online methods  are allowed to access each trace only once. On the other hand, in offline methods, we are able to observe and process the whole data several times. 
	\end{enumerate}

	 The authors in~\cite{gold1978complexity} have proved that the problem of finding the smallest automata based on a given dataset is NP-complete, and later, it was  demonstrated that this is an NP-hard problem~\cite{pitt1993minimum}. These results imply the intractability of passive learning strategies. Active learning methods have also several challenges. Interaction with SUL may be time-consuming or even impossible in particular applications. This is due to the fact that traces need to be independent, meaning that the system requires to be reset for each new input/output trace~\cite{howar2018active}. These are the reasons we have chosen to propose a passive approach to be more practical.
	 
	 Active learning algorithms intrinsically have an online manner, but passive approaches could be online or offline. Online methods are more preferable due to their ability to work in real-time or semi-real-time situations. Moreover, they are generally have less time and memory complexity. They can also be used in incremental synthesis, where each new trace can refine the already synthesized automaton. To the best of our knowledge, our technique is the first passive online approach for synthesizing hybrid automata from input/output traces.
	 
	 In~\cite{medhat2015framework},  a passive offline approach for mining hybrid automata is proposed. The discrete state space is constructed by clustering input-output traces, and the continuous part is modeled by applying statistical methods. In~\cite{soto2019membership}, a passive online algorithm to mine linear hybrid automata is presented.  The idea is to construct a logical formula  corresponding to the synthesis of linear hybrid automata problem, and try to solve that using SMT solver. 
Our research is different in that our goal is to learn automata for non-linear systems, where the continuous behavior of the  system is approximated by an n-degree polynomial equation in each state. Our idea is to use Dynamic Time Warping (DTW) to find similarity between signal segments, and using this measure, we cluster the segments of input/output traces. These clusters make the states of the hybrid automata. To find the continuous behavior of the system, we use polynomial regression method.

The rest of the paper is organized as follows. In Section~\ref{sec:automata}, we provide preliminaries of our research, including hybrid automata and basics of dynamic time warping analysis. In Section~\ref{sec:problem}, we formalize the problem of synthesizing hybrid automata. A motivating example is presented in Section~\ref{sec:example}. Our Passive Online Strategy for Extracting Hybrid Automata based on DTW technique (POSEHAD) is discussed in Section~\ref{sec:alg}. The results are presented in Section~\ref{sec:results} on two case studies of engine-timing hybrid system and the Electronic Control Unite (ECU) of Anti-lock Brake System (ABS). The related work are briefly discussed in Section~\ref{sec:related}. The concluding remarks and future work are presented in Section~\ref{sec:conclusion}.


\section{Preliminaries}
\label{sec:pre}

In this section, we briefly present the preliminary concepts of the paper. The definitions related to the problem input, its output, and the underlying technique of our proposed method are presented in Sections~\ref{sec:trace},~\ref{sec:automata}, and~\ref{sec:warp}, respectively.

\subsection{Trace Modeling}
\label{sec:trace}

As mentioned, our input is a set of input/output traces of the system. Here, we give a brief definition of these traces.

\begin{definition}{Trace of Signal:}
A trace of signal   with the sampling period $c$  is a finite sequence of pairs of timestamps and values $ (t_1,v_1)\cdots(t_p,v_p) $, such that:

\begin{itemize}
	\item $\forall i \in [1,p]:t_i \in \mathbb{R}^{\geq 0}$
	\item $\forall i \in [2,p]: t_i-t_{i-1} = c $ 
	\item $ \forall i \in [1,p]: v_i $ is the value of the signal $S$ at time $t_i$ 
\end{itemize}

\end{definition}

\noindent  As mentioned, our input is a set of $n$ input signals $\{I^1,\cdots,I^n\}$, and $m$ output signals $\{O^1,\cdots,O^m\}$. An input/output trace, represented by $w$ consists of input and output signals.

\begin{definition}{ Input/Output Trace:}
	\label{def:io}
	 An input/output trace is represented by a sequence of tuples:
	\begin{equation}
	\nonumber
	\begin{split}
	 w =  &(t_1,I^1_1,\cdots,I^n_1,O^1_1,\cdots,O^m_1),\cdots,\\
	&(t_p,I^1_p,\cdots,I^n_p,O^1_p,\cdots,O^m_p)\\
	\end{split}
	\end{equation} 
\end{definition}

\noindent , where each tuple consists of a timestamp, and the values of input and output signals at that time.

\begin{definition}{Segmented Input/Output Trace:}
		\label{def:segio}
	A segmented input/output trace $ \psi^{w}_{k,j} $ of a trace $w$ ($ k,j \in [1,p]  ,  k < j $) is a subsequence of $w$:
	\begin{equation}
	\nonumber
	\begin{split}
	 \psi^{w}_{k,j} = 
	 &(t_k,I^1_k,\cdots,I^n_k,O^1_k,\cdots,O^m_k),\cdots,\\
	 &(t_j,I^1_j,\cdots,I^n_j,O^1_j,\cdots,O^m_j)\\
	\end{split}
	\end{equation}

\end{definition}

\subsection{Hybrid Automata}
\label{sec:automata}
The final output of our algorithm is a hybrid automaton. To give an intuition,
consider the automaton of a thermostat system  depicted in Fig.~\ref{therm_hyb}. 
Similar to every state-transition model, a hybrid automaton consists of a set of states and a set of transitions. As an example, the automaton in Fig.~\ref{therm_hyb} has two states, \textit{on} and \textit{off}, and two transitions between them. In a hybrid automaton, there is also a set of {\em continuous variables}. For instance, in the example automaton, the temperature of the environment is a continuous variable  denoted by $x$. A hybrid automaton specifies the valuation of the continuous variables at each state by a predicate. For example, in Fig.~\ref{therm_hyb}, the change rate of $ x $ at state {\em off} is denoted by $ \dot{x} = -0.1x $, where $\dot{x}$ is the first derivation of the variable $ x $ with respect to time. Similarly, the change rate of the variable $ x $ at state {\em on} is specified as $ \dot{x} = 5 - 0.1x $.

\tikzset{every picture/.style={line width=0.75pt}} 

\begin{figure}[htbp]
\centering

\begin{tikzpicture}[scale=0.6, every node/.style={scale=0.6},x=0.75pt,y=0.75pt,yscale=-1,xscale=1]

\draw   (190,173) .. controls (190,139.31) and (217.31,112) .. (251,112) .. controls (284.69,112) and (312,139.31) .. (312,173) .. controls (312,206.69) and (284.69,234) .. (251,234) .. controls (217.31,234) and (190,206.69) .. (190,173) -- cycle ;
\draw   (403,173) .. controls (403,139.31) and (430.31,112) .. (464,112) .. controls (497.69,112) and (525,139.31) .. (525,173) .. controls (525,206.69) and (497.69,234) .. (464,234) .. controls (430.31,234) and (403,206.69) .. (403,173) -- cycle ;
\draw    (111,173) -- (188,173) ;
\draw [shift={(190,173)}, rotate = 180] [color={rgb, 255:red, 0; green, 0; blue, 0 }  ][line width=0.75]    (10.93,-3.29) .. controls (6.95,-1.4) and (3.31,-0.3) .. (0,0) .. controls (3.31,0.3) and (6.95,1.4) .. (10.93,3.29)   ;
\draw    (312,185) -- (401,185.98) ;
\draw [shift={(403,186)}, rotate = 180.63] [color={rgb, 255:red, 0; green, 0; blue, 0 }  ][line width=0.75]    (10.93,-3.29) .. controls (6.95,-1.4) and (3.31,-0.3) .. (0,0) .. controls (3.31,0.3) and (6.95,1.4) .. (10.93,3.29)   ;
\draw    (404,159) -- (314,159) ;
\draw [shift={(312,159)}, rotate = 360] [color={rgb, 255:red, 0; green, 0; blue, 0 }  ][line width=0.75]    (10.93,-3.29) .. controls (6.95,-1.4) and (3.31,-0.3) .. (0,0) .. controls (3.31,0.3) and (6.95,1.4) .. (10.93,3.29)   ;

\draw (248.5,128) node   [align=left] {\begin{minipage}[lt]{17pt}\setlength\topsep{0pt}
off
\end{minipage}};
\draw (464.5,127) node   [align=left] {\begin{minipage}[lt]{17pt}\setlength\topsep{0pt}
on
\end{minipage}};
\draw (262,169.5) node   [align=left] {\begin{minipage}[lt]{62.16pt}\setlength\topsep{0pt}
$ \dot{x}=-0.1x $
\end{minipage}};
\draw (250.5,195.5) node   [align=left] {\begin{minipage}[lt]{31.96pt}\setlength\topsep{0pt}
$ x > 19 $
\end{minipage}};
\draw (134,151) node [anchor=north west][inner sep=0.75pt]   [align=left] {$ x'=20 $};
\draw (358.5,210) node   [align=left] {\begin{minipage}[lt]{31.96pt}\setlength\topsep{0pt}
$ x \le 19 $
\end{minipage}};
\draw (364.5,137.5) node   [align=left] {\begin{minipage}[lt]{40.96pt}\setlength\topsep{0pt}
$ x \ge 21 $
\end{minipage}};
\draw (471,168.5) node   [align=left] {\begin{minipage}[lt]{61.2pt}\setlength\topsep{0pt}
$\dot{x}=5-0.1x$
\end{minipage}};
\draw (462.5,193.5) node   [align=left] {\begin{minipage}[lt]{31.96pt}\setlength\topsep{0pt}
$ x\le22 $
\end{minipage}};

\end{tikzpicture}
\caption{A hybrid automaton modeling a thermostat system}
\label{therm_hyb}
\end{figure}
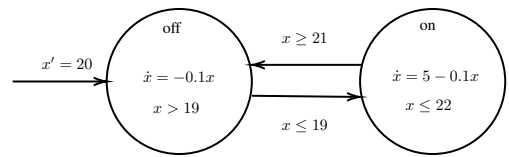

\begin{definition}
 A \textit{hybrid automaton} $H$  is a tuple $\langle X, G , \flow , \jump 
\rangle$, where~\cite{henzinger2000theory}:

\begin{itemize}
	\item $X=\{x_1,...,x_n\}$ is a finite set of real-numbered \textit{variables}.  The number $n$ is called the dimension of $H$.  $\dot{X}= \{\dot{x}_1,...,\dot{x}_n\} $, called the set of dotted variables, represents the first derivatives of variables used for showing their continuous changes. $ X'=\{x'_1,...,x'_n\} $, called the primed variables, represents the values of variables  after the discrete changes.
	\item $G= (V,E)$ is a finite directed multigraph, where the vertices $V$ are called the \textit{control modes}, and  the edges $E$ are called the \textit{control switches}. In Fig.~\ref{therm_hyb}, there are two control modes $ V \in \{\textnormal{on,off}\} $ and two switches $ E \in \{\textnormal{on}\rightarrow \textnormal{off}, \textnormal{off} \rightarrow \textnormal{on}\} $.

	\item $\flow$ is a labeling function $V \rightarrow \pred$ that assigns to each control mode $ v \in V $ a flow condition $ flow(v) $. Each flow condition $flow(v)$ is a predicate whose free variables are from $ X \cup \dot{X} $. In Fig.~\ref{therm_hyb}, the continuous change of $ x $ at mode on is represented by $ \{\dot{x}=5-0.1x\} $
	\item $\jump$ is an edge labeling function $ E \rightarrow \pred$ that assigns to each control switch $ e \in E $ a predicate $\jump(e)$. Each jump condition $ jump(e) $ is a predicate whose free variables are from $ X \cup X'$. Each label can be a jump condition or  a variable assignment. In Fig.~\ref{therm_hyb}, predicate $ x > 21 $ is a jump condition on the control switch $\textnormal{on} \rightarrow \textnormal{off}$ and $ x'=20 $ is an assignment to the variable $x$. Note that control modes that are the destination of  control switches without a source control mode are the initial control modes, and the assignments of these control switches represent the initial values of the variables.
	
\end{itemize}

\end{definition}

\subsection{Dynamic Time Warping Analysis}
\label{sec:warp}

Our synthesis algorithm is based on 
Dynamic Time Warping (DTW).  It was first introduced in~\cite{bellman1959adaptive} to detect similarity between two nonlinear time series. The technique has been widely used in several areas, such as speech, handwriting and gesture recognition, signal processing, data mining, and time series clustering. DTW employs dynamic programming to calculate the optimal match between two given time series $ X=(x_1,...,x_N) $ and $ Y = (y_1,...,y_M) $ with the both time and storage complexity of $ O(N M) $~\cite{senin2008dynamic}.  As an example, consider two time series $X$ and $Y$ that are depicted in the right plot of Fig.~\ref{fig3}. DTW tries to find the most similar corresponding data points between the two time series. The left plot in Fig.~\ref{fig3} demonstrates the alignment matrix of time series $X$ and $Y$. This matrix indicates the corresponding data points in time series $X$ and $Y$. The diagonality in the emerging path shows the similarity between the two time series. 

Given two sequences $ X $ and $ Y $, the algorithm constructs a distance matrix $C$ for their alignment:

\begin{equation}
	C \in \mathbb{R}^{N\times M}:c_{ij}=||x_i-y_j||, i \in [1:N], j \in [1:M]
\end{equation}

\noindent DTW builds an alignment path $ P = (p_1,p_2,...,p_k)$, such that $ \forall l \in [1,k]: p_l=(i,j) : i \in [1:N] \;,\; j \in [1:M]$,
and finally it finds a path with minimum distance from $ p_1=(1,1) $ to $ p_k = (N,M) $. The optimal distance of time series $ X $ and $ Y $ is calculated by Eq.~\ref{eq3}.

\begin{equation}
	\label{eq3}
	dist^{DTW}(X,Y) = \sum_{l=1}^{k}c_{i_{p_l}j_{p_l}}
\end{equation}

DTW analysis not only calculates the minimum distance between two time series, but it also determines the corresponding data points in the two time series.
We denote the first elements of path $ P $ as $P_i= (i_{p_1},i_{p_2},...,i_{p_k}) $ and the second elements of path $ P $ as $ P_j= (j_{p_1},j_{p_2},...,j_{p_k}) $. Then we could measure the diagonality of path $ P $ as follows:

\begin{equation}
	\label{eq4}
	diag^{DTW}(X,Y) = correlation(P_i,P_j)
\end{equation}


\section{Problem Statement}
\label{sec:problem}
Based on the previous definitions, the problem of mining hybrid automaton is formally defined as follow:
\begin{center}
	\fbox{\rule{1mm}{0mm}
		\begin{minipage}[t]{.95\columnwidth}
			\label{prb:repair}
			
		\textbf{Given} is a set of training input/output traces $\Phi^r= \{w_{r1},w_{r2},\cdots,w_{rn}\} $ and a set of test input/output traces $\Phi^t= \{w_{t1},w_{t2},...,w_{tm}\} $ of a black-box system.
		
			Our goal is to design a  passive online algorithm that synthesizes a hybrid automaton $H$  modeling the behavior of the system based on $\Phi_r$. The precision of the synthesized automaton is evaluated by the following cost function:
			
			\begin{equation}
			\label{costfunc}
			cost = \frac{1}{|\Phi^t|} \sum_{i=1}^{|\Phi^t|} || H(w^{I}_{ti})- w^{O}_{ti}||
			\end{equation}  
	\noindent	, where $ H(w^{I}_{ti}) $ denotes the resulting output signals, when the input signals of trace $ w_{ti}$ ($ w^I_{ti} $)  are given to the automaton $H$. 
		\end{minipage}
	}
\end{center}

\section{Motivating Example}
\label{sec:example}
Consider an engine timing system, having three input signals: (1) input throttle, which is the desired speed of the system in Rounds Per Minute (RPM), (2) load torque in joules per radian, and (3) time as an independent input signal. It  has one output signal, called speed engine in RPM. Our goal is to model the behavior of the speed engine based on the system's input signals using a hybrid automaton. One  input/output trace of this system is illustrated in Fig.~\ref{fig1}.

\section{POSEHAD Algorithm }
\label{sec:alg}
In this section, we present a Passive Online Strategy for Extracting Hybrid Automata based on DTW (POSEHAD). POSEHAD consists of the following steps:

\begin{enumerate}
	\item Signal segmentation:  Traces are processed one by one due to the online manner of this approach. The first step of processing each new input/output trace $w$ is to segment it by a change point detection algorithm, as discussed in Section~\ref{sigseg}. As a result, we have a set of segmented input/output traces corresponding to $w$.
	
	\item Signal similarity detection: Having a set of segmented input/output traces corresponding to $ w$, we need a similarity detection criterion to categorize the similar segments into a set of states. This step is discussed in Section~\ref{simdet}.
	
	\item Extracting the discrete state space: So far, we have a set of segments of $w$ and a similarity detection index. The next step is to mine the discrete parts of automaton by identifying its states and transitions. The algorithm takes each segment and calculates the similarity between this segment and the existing set of states. If there are  states which similarity with the segment exceeds a predefined threshold, then (1) the state with maximum similarity will be chosen, (2) the segment will be appended to that state, and (3) the transitions will be updated. If no similar state is found, then (1) a new state will be created, (2) the segment will be added to that state, and (3) the set of transitions will be updated. The details of this step are presented in Section~\ref{extdisc}
	
	\item Mining Jump Conditions: We utilize the updated sets of states and transitions to mine the jump conditions of hybrid automaton. This step is discussed in Section~\ref{sec:jump}.
	
	\item Discovering the flow conditions: The algorithm takes segments of each state and extracts the flow conditions, as discussed in Section~\ref{discinv}.
\end{enumerate}

%
%


	\subsection{Signal segmentation}
	\label{sigseg}
The first preprocessing step of our technique is signal segmentation, where we divide each input/output trace into multiple segments. There are different approaches for signal segmentation, each proposed for a specific type of signal. Note that we assume that the SUL has synchronous behavior,  meaning that each change in inputs affects the outputs with a negligible delay. Our goal is to find a set of change points, where each point demonstrates a drastic change in at least one input or output trace. In this work, we utilize window-sliding change-point detection algorithm~\cite{truong2020selective} to perform segmentation. This method considers an immediate past and an immediate future window with size $W$ for each data point in a signal with $N$ data points. It then computes the discrepancy of each data point based on these two windows and finally considers data points with maximum discrepancy as a change point in a signal. Eq.~\ref{eq1} takes an input/output trace $w$  and returns a set of change-points belonging to $w$ denoted by $CP_w $.

	\begin{equation}
	\label{eq1}
	CP_w = window\_sliding\_algorithm(w)
	\end{equation}

	Using $CP_w$, the input/output trace is divided into a set of segmented traces. All the segmented input/output traces related to $w$ (as defined in Definition~\ref{def:segio}) are denoted by $ \Psi^{w} $, as shown in Eq.~\ref{eq2}.
	
	\begin{equation}
	\label{eq2}
	\begin{split}
	&\Psi^{w} = \{\psi^{w}_{1,i},\psi^{w}_{i+1,j},\cdots,\psi^{w}_{k,l-1},\psi^{w}_{l,p}\} 
	\end{split}
	\end{equation} 
	
\noindent For each change point, a neighborhood is defined as follows. This neighborhood is used to find the transition jumps, as discussed in Section~\ref{sec:jump}.

\begin{definition} \label{def:cp} {Change-Point Neighborhood}: For a change-point $cp$ in signal  $w=  (t_1,I^1_1,\cdots,I^n_1,O^1_1,\cdots,O^m_1),\cdots,\\(t_p,I^1_p,\cdots,I^n_p,O^1_p,\cdots,O^m_p)$, the change-point neighborhood of  $ cp $  is a vicinity of $ v $ data-points, represented by a finite sequence of tuples:
\begin{equation}
\begin{split}
 \psi^{w}_{cp} =  &(t_{cp-v},I^1_{cp-k},\cdots,I^n_{cp-k},O^1_{cp-k}\cdots,O^m_{cp-k}),\cdots,\\
&(t_{cp},I^1_{cp},\cdots,I^n_{cp},O^1_{cp-k}\cdots,O^m_{cp}),\cdots,\\
&(t_{cp+v},I^1_{cp+k},\cdots,I^n_{cp+k},O^1_{cp+k}\cdots,O^m_{cp+k})\\
\end{split}
\end{equation} 
\end{definition}
	
	\begin{figure}[htbp]
		\centerline{\includegraphics[width=0.9\columnwidth]{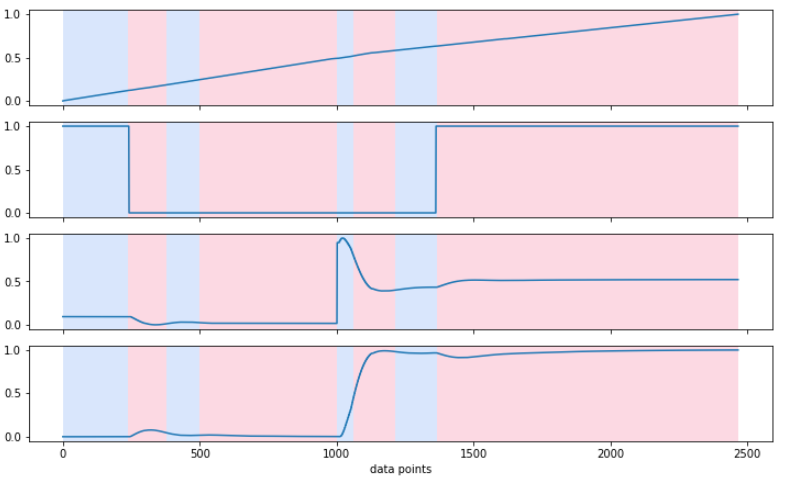}}
		\caption{Segmentation of an input/output trace $ w$ of the engine timing system based on window-sliding algorithm}
		\label{fig1}
	\end{figure}
	Fig.~\ref{fig1} illustrates $ \Psi^{ w } $ of one input/output trace ($w$) of the engine-timing case study. The topmost  trace is the working time considered as an input signal. The second sampled trace shows the input throttle. The third trace is the input load torque, and the last trace is the output engine speed. Note that all the input and output traces of a sample will participate in segmentation, and each change point demonstrates a drastic change  in at least one  trace.
	
\subsection{Signal similarity detection}
\label{simdet}
After segmentation, we need a mechanism to find a similarity index between segmented input/output traces with non-linear behavior, so that we can categorize them. There are three major challenges in this regard:

	\begin{enumerate}
	\item \textbf{Unequal lengths of segments}:
	Different criteria, such as euclidean distance and correlation coefficient~\cite{kianimajd2017comparison} are suggested to find similarity between two time series. However, these measures can only evaluate the similarity between two equal time series, and hence, we cannot use them due to the  unequal length of segments.
	
	\item \textbf{Online clustering of segments}: In~\cite{medhat2015framework},  statistical measures, such as the average slope or the mean of each segment are used  to cluster segments. Our constraint in using this approach is that in an online manner,  we do not have access to all signals at the time of clustering. At each point, we have a clustering for signals that we have seen so far, and each new trace will be added to the clustering, once it is received and processed.
	
	\item \textbf{Nonlinearity of signals}:
	In~\cite{soto2019membership}, the similarity between segments is calculated based on the coefficients of linear segments. This is based on the assumption  that all segments have a linear behavior, and segments with equal slopes (with a degree of freedom) will be placed in one state of the synthesized linear hybrid automaton. For non-linear behaviors (that we model with quadratic polynomials), we could not use equation's coefficients because of non-orthogonal basis appeared in quadratic or higher order polynomials.
\end{enumerate}

Our method in this paper for signal similarity detection is based on Dynamic Time Warping (DTW). We have selected this technique for the following reasons:
	\begin{enumerate}
	\item This method can find the similarity between two time series with unequal lengths. 
	\item The algorithm intrinsically finds the geometrical similarities between two time series that could not be done based on equation's coefficients.
	\end{enumerate}

We use  both the distance  $ dist^{DTW} $ and the diagonality rate $ diag^{DTW} $  to define a similarity index between two segmented traces, $\psi$ and $\psi'$.

	\begin{equation}
\nonumber
sim\_index(\psi,\psi') = (dist^{DTW}(\psi,\psi'),diag^{DTW}(\psi,\psi'))
\end{equation}

\begin{figure}[htbp]
	\centerline{\includegraphics[width=1.0\columnwidth]{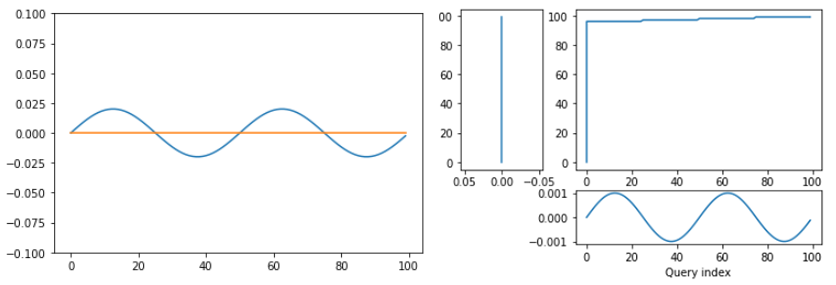}}
	\caption{An Example of two dissimilar signals with low distance and low diagonalty}
	\label{fig14}
\end{figure}

\begin{figure}[htbp]
	\centerline{\includegraphics[width=1.0\columnwidth]{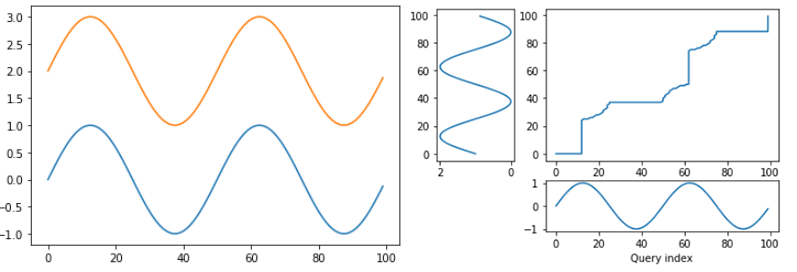}}
	\caption{An Example of two dissimilar signals with high distance and high diagonalty}
	\label{fig15}
\end{figure}

Both of these criteria are necessary for finding the similarity between two signals. To clarify, consider two time series with small ranges as  depicted in Fig.~\ref{fig14}. The distance measure alone can not distinguish their difference, since the distance can be small due to the small ranges. On the other hand, based on Fig.~\ref{fig15}, if we only consider the diagonallity rate as a metric, we will not be able to discriminate between two equal time series with different ranges.

\begin{figure}[htbp]
	\centerline{\includegraphics[width=1.0\columnwidth]{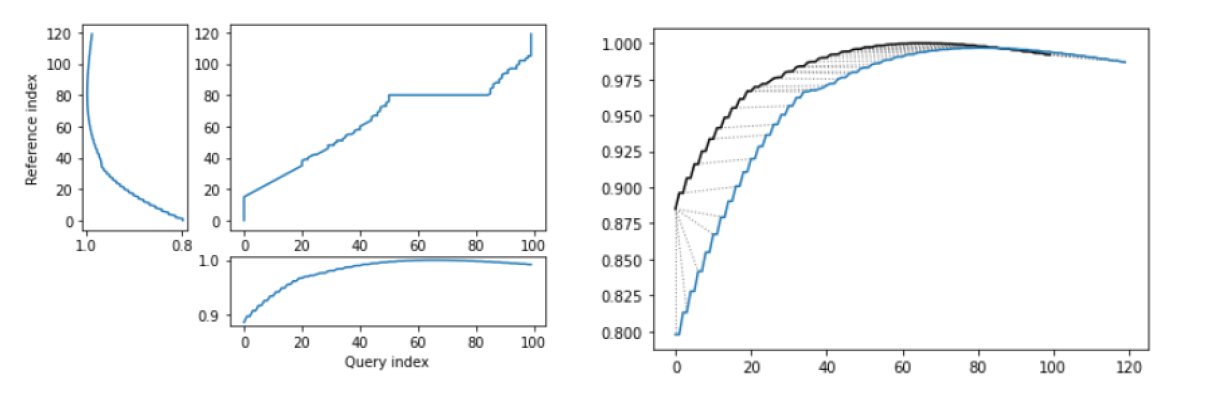}}
	\caption{Left figure: alignment matrix of two similar signals, Right figure: corresponding points of the two signals}
	\label{fig3}
\end{figure}

\begin{figure}[htbp]
	\centerline{\includegraphics[width=1.0\columnwidth]{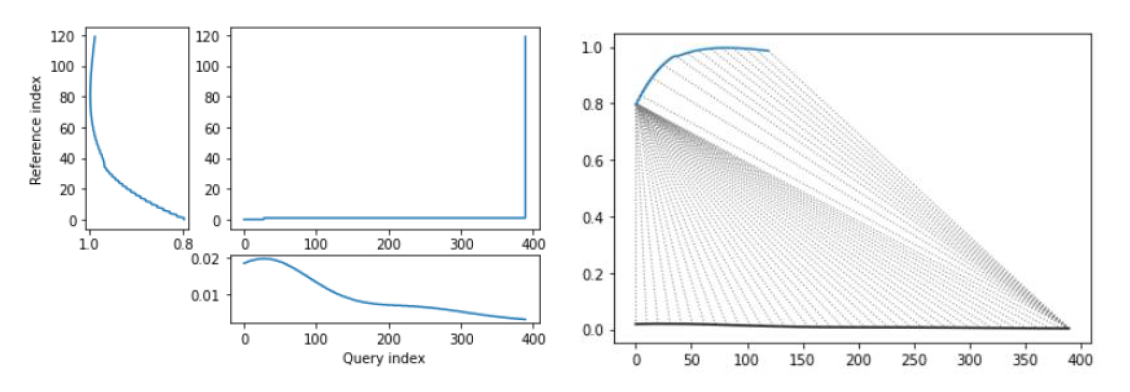}}
	\caption{Left figure: alignment matrix of two dissimilar signals, Right figure: corresponding points of the two signals}
	\label{fig4}
\end{figure}

Fig.~\ref{fig3} shows an example of two similar segments. The similarity can be inferred from the small distance of dotted lines in the right figure, and the near-diagonal path line illustrated in the left figure. Fig.~\ref{fig4} depicts two dissimilar segmented traces. The non-diagonal shape of the path line in the left figure demonstrates that there is no one-to-one correlation between the two signals.


\begin{algorithm}
	\caption{POSEHAD}
	\label{alg1}
	\begin{algorithmic}[1]
		\STATE{\textbf{Inputs:} $segmented\_IO\_traces(\Psi^{w})$ , \\
			$  distance\_threshold(\theta_{dis}) $,$ diagonality\_threshold(\theta_{diag}) $,  $ states $,  $ transitions $ }
		\STATE {$ current\_state \leftarrow null $} \label{line:emptycurrent}	
		\FOR{ each $ seg $ in $ \Psi^{w} $}\label{line:foreachseg}
		\STATE $ tran\_seg \leftarrow changepoint\_neighborhood(seg[0]) $ \label{line:cpneighborhood}
		
		\IF{$states$ is empty} \label{line:ifempty}
		\STATE $new\_state \leftarrow create\_new\_state()$
		\STATE $states \leftarrow states \cup new\_state$
		\STATE $ state\_segments(new\_state) \leftarrow \{seg\}$
		\STATE $new\_tran \leftarrow (current\_state,new\_state)$
		\STATE $transitions \leftarrow transitions \cup new\_tran$
		\STATE $ tran\_segments(new\_tran) \leftarrow \{tran\_seg\}$
		\STATE $ update\_transition\_confidence\_level() $ 
		\STATE $ current\_state \leftarrow new\_state $  \label{line:endifempty}
		
		\ELSE \label{line:elseempty}
		\STATE $ candidate\_state \leftarrow null$ \label{line:initcand}
		\STATE $ candidate\_dist \leftarrow +\infty$
		\STATE $ candidate\_diag \leftarrow 0$\label{line:endinitcand}
		
		\FOR{ each $ state $ in $ states $} \label{line:sim}	
		\IF{$ dist(seg,state) < candidate\_dist$ and $diag(seg,state) > candidate\_diag $} 	
		\STATE $ candidate\_dist  \leftarrow dist(seg,state) $
		\STATE $ candidate\_diag  \leftarrow diag(seg,state) $
		\STATE $candidate\_state \leftarrow state$
		\ENDIF
		\ENDFOR	\label{line:endsim}
		
		\IF{$ candidate\_dist < \theta_{dis}$ and $candidate\_diag > \theta_{diag}$ }\label{line:hascand}
		\STATE $ state\_segments(candidate\_state) \leftarrow \{seg\}$
		\STATE $new\_tran \leftarrow (current\_state,candidate\_state)$
		\STATE $transitions \leftarrow transitions \cup new\_tran$
		\STATE $ tran\_segments(new\_tran) \leftarrow$ \\ ~~~~~~~~$tran\_segments(new\_tran) \cup \{tran\_seg\}$
		\STATE $ update\_transition\_confidence\_level() $ 
		\STATE $ current\_state \leftarrow candidate\_state $  \label{line:endhascand} 
		
		\ELSE \label{line:newstate}
		\STATE $new\_state \leftarrow create\_new\_state()$
		\STATE $ state\_segments(new\_state) \leftarrow \{seg\}$
		\STATE $new\_tran \leftarrow (current\_state,new\_state)$
		\STATE $transitions \leftarrow transitions \cup new\_tran$
		\STATE $ tran\_segments(new\_tran) \leftarrow \{tran\_seg\}$
		\STATE $ update\_transition\_confidence\_level() $ 
		\STATE $ current\_state \leftarrow new\_state $ 
		\ENDIF \label{line:endnewstate}
		\ENDIF
		\ENDFOR
		\FOR{ each $ tran $ in $ transitions $}
		\STATE $update\_jump(tran,tran\_segments
		(tran),$\\
		$~~~~~~~~~~~~~~~~~~state\_segments(src(tran)))$ \label{line:jump}
		\ENDFOR
		\FOR{ each $ state $ in $ states $}
		\STATE $ update\_flow(state,state\_segments(state)) $\label{line:flow}
		\ENDFOR
		\STATE {\textbf{Outputs:} $states$, $transitions$}
	\end{algorithmic}
\end{algorithm}

 \begin{figure}[htbp]
	\centerline{\includegraphics[width=1.0\columnwidth]{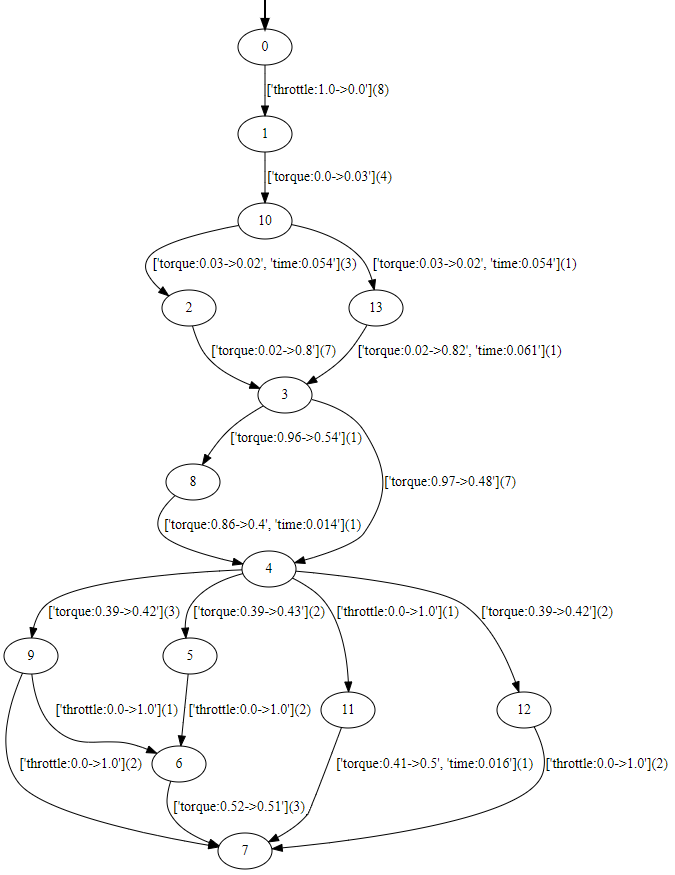}}
	\caption{The synthesized hybrid automaton for the engine timing system}
	\label{fig5}
\end{figure}

\subsection{Extracting the discrete state space}
\label{extdisc}
 Algorithm~\ref{alg1} exhibits the procedure of mining hybrid automaton. The algorithm runs for each new unprocessed input/output trace $w$. After performing segmentation on $w$, the segmented traces $\Psi^{w}$ are given to the algorithm. The other inputs to the algorithm are  distance threshold ($ \theta_{dist} $), diagonality threshold ($ \theta_{diag} $), and the states and transitions  of the automaton that have been synthesized so far. The algorithm returns the updated set of states and transitions as output. For the first sample, we pass an empty set of states  and transitions to the algorithm. 

  The set of states is a dictionary, where the keys are state names, and the value of each key  $ state $ is a  set of similar segments, called \textit{state segments}  that are assigned to that state. The set of transitions is a dictionary, where each key contains  two states (determining the source and the target of the transition), and each value is a set of change-point neighborhoods (as defined in Definition~\ref{def:cp}) corresponding to that transition. This set is called \textit{transition segments} and it  demonstrates the behavior of input signals during traversal between two states.
  
 In line~\ref{line:emptycurrent} of Algorithm~\ref{alg1}, at the beginning, the current state is assigned to null.
  In line~\ref{line:foreachseg}, the algorithm iterates over each segment of the trace $w$. In line~\ref{line:cpneighborhood}, we calculate the change-point neighborhood of the first element in the segment $ seg $ ($seg[0]$). Note that $seg[0]$ is actually the change-point of the segment $seg$, and the function $changepoint\_neighborhood$ returns the neighborhood of the given change-point. The change-point neighborhood will later be assigned to the transition segments of the corresponding transition. In lines~\ref{line:ifempty}-\ref{line:endifempty}, if the set of states is empty, a new state is created, and the current segment will be added to it. The set of transitions, and the transitions' confidence level will also be updated. For each transition, we assign a probability to each input signal called confidence level. It determines how much that signal is responsible for triggering the transition. The details will be discussed in Section~\ref{sec:jump}.
  In lines~\ref{line:initcand}-\ref{line:endinitcand}, a candidate state will be initialized by infinite distance and zero diagonality rate.
In lines~\ref{line:sim}-\ref{line:endsim}, if the set of states is not empty, the algorithm iterates over each state and calculates the distance and diagonality between the current segment and the state. If the distance of the current state is less than $ candidate\_dist $ and the diagonality rate of the current state is more than $ candidate\_diag $, this state will be selected as the candidate state.

 In lines~\ref{line:hascand}-\ref{line:endhascand}, the algorithm checks if the candidate state passes the distance and diagonality thresholds. If so, the current segment will be added to the state segments of that state. Accordingly, the set of transitions will be updated, and the transition from the current state to the candidate state will be added to it. In lines~\ref{line:newstate}-\ref{line:endnewstate}, if the current segment does not match with any state, then it will be added to a new created state. The set of transitions and the transition's confidence level will also be updated.

As an example, for the engine timing hybrid system,  we fed 10 input/output traces to our procedure. For each trace, we segmented it, and gave the segments, along with our specified thresholds to Algorithm.~\ref{alg1}. At the end, the set of states and transitions are used to construct the discrete part of an automaton, as depicted in Fig.~\ref{fig5}.  As mentioned, for each state, there is a set of segments, called state segments. Consider state '7' in Fig.~\ref{fig5}, for example. There are  8 state segments for this state, as depicted in Fig.~\ref{fig6}. In this figure, each column corresponds to one segment belonging to state '7'. Note that each segment consists of all input and output traces. Similarly, for each transition $ tran $, we have a set of equal-length change point neighborhoods, called transition segments. They are utilized to detect which input signal is responsible for that transition. For example, there are 8 transition segments corresponding to the transition $ 0 \rightarrow 1 $, which are illustrated in Fig.~\ref{fig7}. Each column depicts the input/output trace of one transition segment.

\subsection{Mining jump conditions}
\label{sec:jump}
In line~\ref{line:jump} of Algorithm~\ref{alg1}, the jumps of synthesized transitions are mined using the transition segments and state segments. The cause of each jump may be due to the behavior of the input signals, which we call the input event jump, and/or due to the system staying in a state for a certain amount of time, called time condition. These categories do not form a mutually exclusive dichotomy. We discuss mining these two types of transition jumps in Sections~\ref{sec:inputevent} and~\ref{sec:timecond}.
\subsubsection{Input events}
\label{sec:inputevent}
 When a transition is created between two states, a confidence level for each input signal is assigned, which indicates how much that signal is responsible for triggering the transition. Initially, the confidence level of all input signals of a transition are set as one. For each new transition segment added to a transition set, we update the confidence level of each signal based on the euclidean distance between new change-point neighborhood and the current transition segments. The less similar an input  trace of the new transition segment to all the previous corresponding  traces in the transition segments, the less confidence level will be assigned to that input signal. Note that we are able to utilize euclidean distance due to the equal length of transition segments.

Consider transition $ 0 \rightarrow 1 $ of the automaton in Fig.~\ref{fig5} as an example. The transition segments belonging to this transition are depicted in Fig.~\ref{fig7}. Note that we only consider the input signals (not the output), since only input signals can trigger transitions. When the first sample is fed to the algorithm, the transition segments of this transition consists of only the first column of Fig.~\ref{fig7} and the confidence level of this transition is assigned to $ \{1.0,1.0,1.0\} $ for the input signals of time, input throttle, and input torque, respectively. After feeding all input/output traces, there are 8 change-point neighborhoods added to this transition segments set, and the final confidence level for this transition is updated to $ \{0.168, 1.0   , 0.985\} $. These values indicates that the input throttle (the second input signal) is the most effective one for transition $ 0 \rightarrow 1 $. Intuitively, when we look at Fig.~\ref{fig7}, it could be seen that in different samples, this transition happens in different time and torque values based on the first and third row respectively, but when this transition happens, the input throttle always has the same manner based on the second row.

During the update\_transitions\_set phase in Algorithm~\ref{alg1}, these confidence levels will be gradually updated with new input/output traces. When we want to represent our automaton using the output of Algorithm~\ref{alg1}, we just need to select an input signal with the maximum confidence level for each transition $ a \rightarrow b $ to model the transition's jump condition. It is also worth mentioning that since we are modeling deterministic systems, if there are more than one transitions going out of a state with overlapping conditions, we distinguish them by adding more input signals to the conditions (making ``and'' conditions).

\subsubsection{Time conditions}
\label{sec:timecond}
The intuitive meaning of a time condition is that the transition is triggered, when the system stays in a state for a specific amount of time.
 For mining these conditions, the elapsing time of the system being in the source state of the transition is calculated according to state segments. Then the time causality of a transition is found by computing the variance of state segments durations. For instance, consider transition $ 10 \rightarrow 13 $ in Fig.~\ref{fig5}. It could be inferred from the jump condition 'time:0.054' that if the system remains in state '10' for a period of '0.054'(based on time normalization between zero and one), the transition $ 10 \rightarrow 13 $ will be triggered because of this time condition. 

\subsection{Mining the flow conditions}
\label{discinv}
In line~\ref{line:flow} of Algorithm~\ref{alg1}, the flow conditions of the synthesized states are mined.
 As mentioned earlier, we approximate the continuous state of a nonlinear hybrid automaton by an n-degree polynomial. We employ the polynomial regression method to estimate the parameters of that equation, based on the state segments of each state. For extracting the flow conditions, the partial derivative of inputs is calculated from the polynomial equation.
 
\begin{figure}[htbp]
	\centerline{\includegraphics[width=1.0\columnwidth]{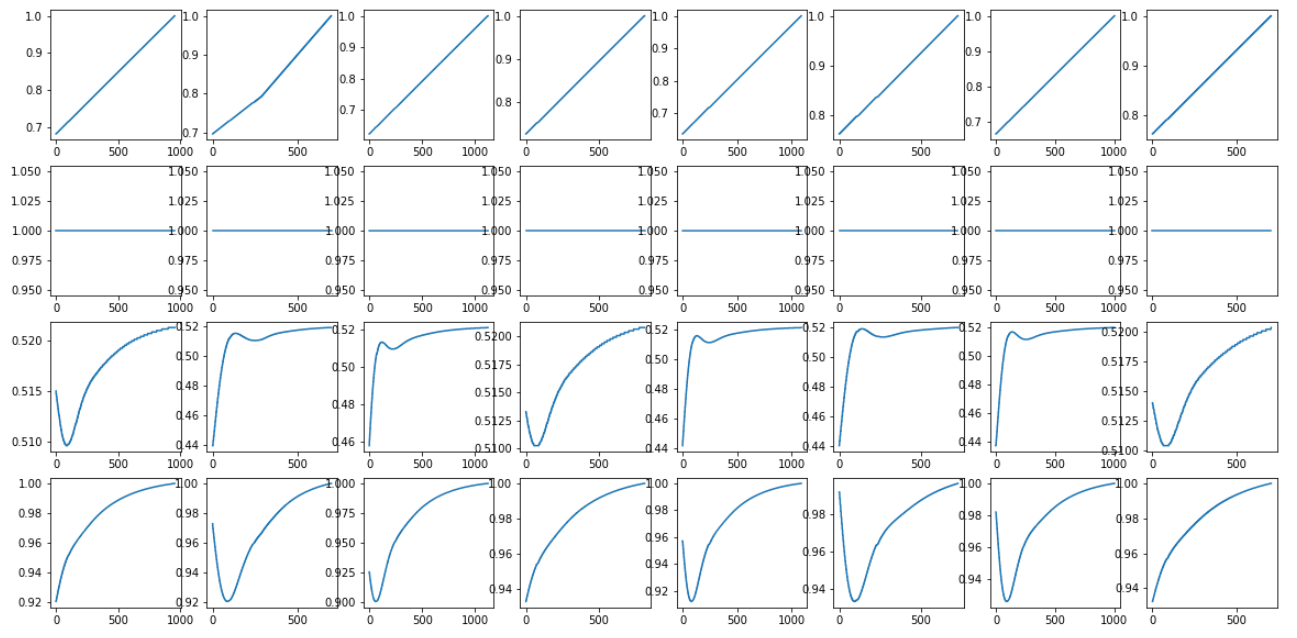}}
	\caption{State segments belonging to state '7'}
	\label{fig6}
\end{figure}

\begin{figure}[htbp]
	\centerline{\includegraphics[width=1.0\columnwidth]{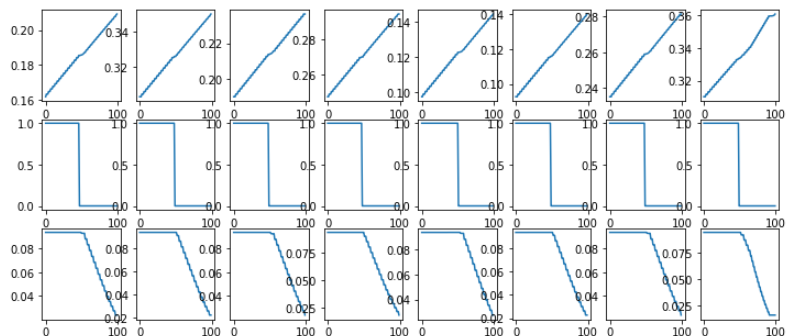}}
	\caption{Change-point neighborhoods belonging to transition '0-1'}
	\label{fig7}
\end{figure}

Consider state '7' of the automaton in Fig~\ref{fig5} as an example. All the state segments corresponding to this state are presented in Fig.~\ref{fig6}. The algorithm applies polynomial regression to find the equation of the output signal based on the input signals:

\begin{equation}
\begin{split}
y =& 0.01-0.32x_0-0.0x_1-10.91x_2-0.84x_0^2-0.32x_0 x_1 \\ 
&+4.3x_0 + x_2+0.0x_1^2-10.91x_1x_2+19.02x_2^2\\ 
\end{split}
\end{equation}

\begin{equation}
\frac{\partial y}{\partial x_0} = -0.32-1.68x_0-0.32 x_1+4.3 x_2 
\end{equation}

\begin{equation}
 \frac{\partial y}{\partial x_1} = -0.32x_0 -10.91x_2 
\end{equation}

\begin{equation}
\frac{\partial y}{\partial x_2} = -10.91 +4.3x_0 - 10.91x_1+38.04x_2
\end{equation}

\section{Experimental Results}
\label{sec:results}

We evaluate our proposed framework on two case studies. The first one  is our example of  engine timing system, which is available as a built-in hybrid model in the Simulink toolbox. Our second case study is the Electronic Control Unit (ECU) of an Anti-lock Brake System (ABS), which data is taken from the automobile industry. We have employed the sliding window change-point detection technique implemented in rupture python library~\cite{truong2020selective} to perform signal segmentation. 

\subsection{Engine timing system}

To generate the traces of our example of engine timing system, we utilized the built-in hybrid model in the Simulink toolbox.
To evaluate our algorithm, we generated 10 training input/output traces, trying  to cover the whole state space of system by different permutations of input signals. We fed input/output traces to Algorithm~\ref{alg1} one after another. The final synthesized hybrid automaton is depicted in Fig.~\ref{fig5}.

\begin{figure}[htbp]
	\centerline{\includegraphics[width=0.8\columnwidth]{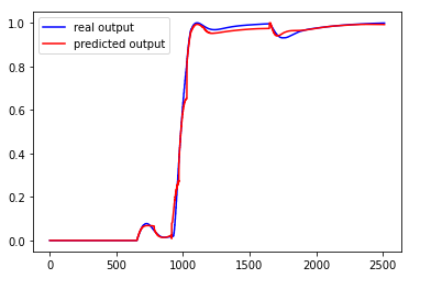}}
	\caption{Predicted and real outputs of an engine timing for a test trace }
	\label{fig8}
\end{figure}

To evaluate the synthesized automaton, we fed a set of test input/output traces to the synthesized hybrid automaton and compared the predicted outputs with the real outputs. The  value of the cost function introduced in Eq.~\ref{costfunc} using the set of test input/output traces was equal to $ "0.0266" $. Predicted and real outputs  for a test trace are demonstrated in Fig.~\ref{fig8}, which shows a high precision of the synthesized automaton.

\subsection{ECU of Anti-Lock Brake System}
We selected the ECU of ABS as our second case study to learn its behavior and evaluate the POSEHAD algorithm. ABS aims to prevent the car from slipping in adverse weather conditions when the driver pushes the braking pedal. ECU receives each wheel's velocity and the brake signal at each moment and then controls the pressure applied to each hydraulic valves in the Hydraulic Control Unit (HCU) of ABS.

\begin{figure}[htbp]
	\centerline{\includegraphics[width=0.8\columnwidth]{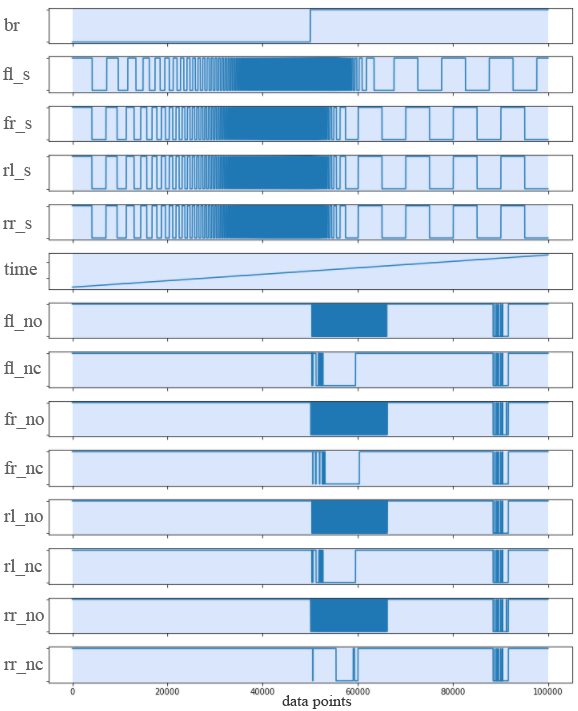}}
	\caption{Input/output signals of ECU}
	\label{fig9}
\end{figure}

As presented in Fig.~\ref{fig9}, the ECU consists of brake (br), front left wheel speed (fl\_s), front right wheel speed (fr\_speed), rear left wheel speed (rl\_s), rear right wheel speed (rl\_s), and time as input signals. For each wheel, it has a normally open valve and a normally close valve, as output signals. For instance, rear left normally open (rl\_no) and rear left normally close (rl\_nc) belong to the rear left wheel.

\begin{figure}[htbp]
	\centerline{\includegraphics[width=0.9\columnwidth]{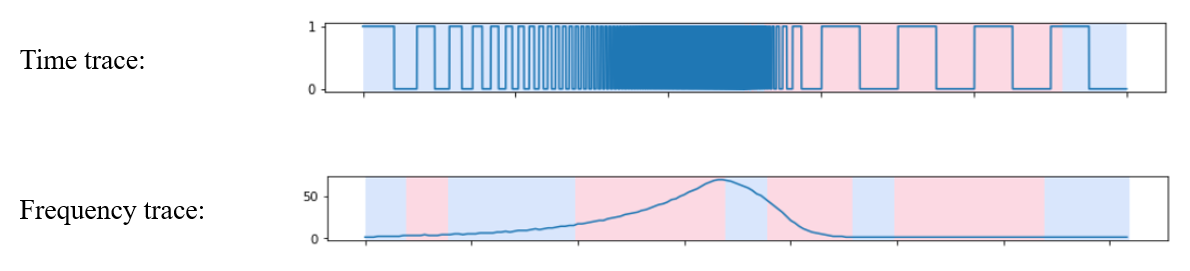}}
	\caption{Converting a trace of (time,value) to a  trace of (time,frequency) }
	\label{fig10}
\end{figure}

Each trace in this case study consists of 100,000 data points taking a value of zero or one to construct a square wave with variable frequency. Due to the microscopic granularity of these traces, we applied a preprocessing phase to convert each trace from time-value to time-frequency as presented in Fig.~\ref{fig10}.

\begin{figure}[htbp]
	\centerline{\includegraphics[width=0.9\columnwidth]{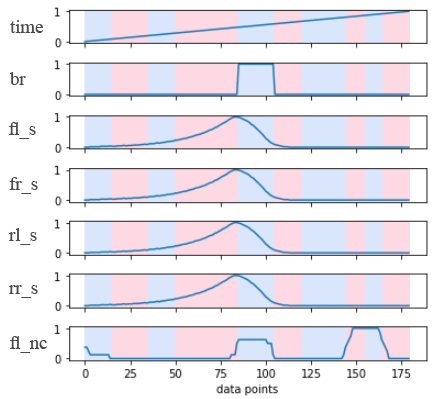}}
	\caption{One Input/output trace for front left normally close (fl\_nc)}
	\label{fig11}
\end{figure}

For each output signal, a hybrid automaton is learned using the POSEHAD algorithm.  Fig.~\ref{fig11} presents an example of input/output traces in the training set of fl\_nc.

\begin{figure}[htbp]
	\centerline{\includegraphics[width=0.8\columnwidth]{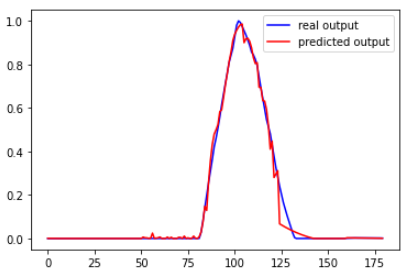}}
	\caption{Real and predicted values of front left normally open valve }
	\label{fig12}
\end{figure}

We fed 8 input/output traces to the POSEHAD algorithm to learn the behavior of the front left normally open valve. The real  and the predicted values of an input/output test signal are depicted in Fig.~\ref{fig12}. Based on a set of test input/output traces and Eq.~\ref{costfunc}, the cost of the predicted outputs was equal to $ "0.247" $.  Finally, we performed a reverse transformation from pairs of (time, frequency) to the pairs of (time, value) to have output traces in the domain of input signals.

\begin{figure}[htbp]
	\centerline{\includegraphics[width=0.9\columnwidth]{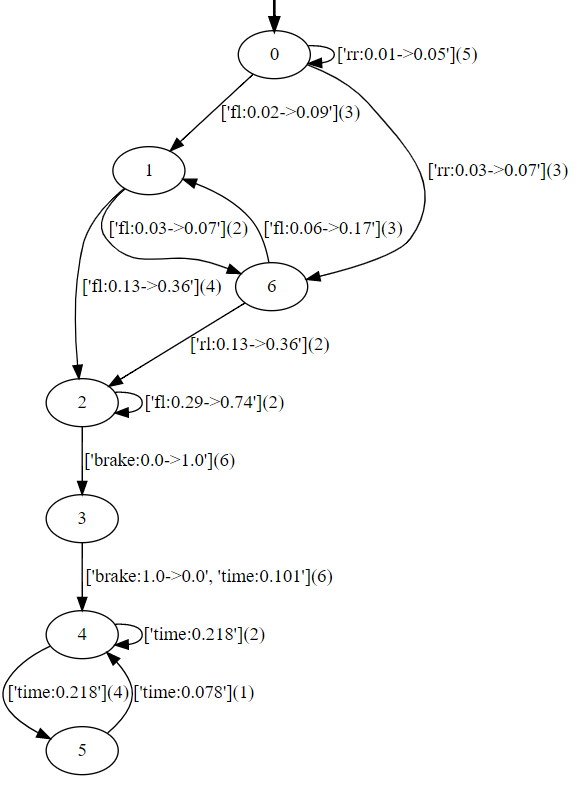}}
	\caption{The synthesized hybrid automaton for the front left normally open output signal}
	\label{fig13}
\end{figure}

The final synthesized hybrid automaton that has modeled the behavior of the front left normally open valve is illustrated in Fig.~\ref{fig13}.
 The label of each transition denotes the jump condition, along with the number of training traces taking that transition. For example, label ['brake:0.0-$>$1.0'](6) assigned to transition $ 2 \rightarrow 3 $  demonstrates that there are 6 input/output traces in the training sample that move from state 2 to state 3. The condition of this transition is changing the brake input signal from zero to one.

\section{Related Work}
\label{sec:related}

Specification mining has been extensively studied in several papers~\cite{le2018deep,kang2021adversarial,zhang2020temporal}. In~\cite{le2018deep}, the authors proposed a deep specification miner for extracting Finite State Automata (FSA) models. They used each execution trace to construct a prefix tree acceptor (PTA), and merge PTAs of different traces to synthesize a number of FSAs.  Our work is different in that we deal with mining the specifications of embedded control systems with both continuous and discrete variables.

Another line of work related to this research
is system identification~\cite{juloski2005comparison,lauer2019hybrid,blavzivc2020hybrid,pillonetto2016new}. These works attempt to model a system with a set of differential equations and estimate their parameters to identify hybrid system's behavior. They tackle the problem from the control system perspective.  The main challenge of these approaches is that the model complexity needs to be known~\cite{pillonetto2016new}. 

Automata learning is a field of studies with a well-established and extensive research history~\cite{okudono2020weighted,howar2018active,maier2014online,weiss2018extracting}.
The closest articles to our work are those that model the behavior of hybrid systems using a  hybrid automaton~\cite{medhat2015framework,lamrani2018hymn,soto2019membership}.
In~\cite{medhat2015framework}, the authors proposed a passive offline framework for learning hybrid automata. Compared to their approach, our algorithm is online, meaning that it could refine the synthesized hybrid automata by new input/output traces. The other point is that their method for mining the discrete state space requires a large amount of training data. They also modeled the continuous state space of the system by linear equations. In contrast, our method could approximate the non-linear behaviors with an n-degree polynomial equation.
The authors of~\cite{soto2019membership} proposed a passive online  algorithm to mine linear hybrid automata. They assumed that each trace could be segmented into a set of piecewise linear functions (PLFs). PLFs with similar slopes are then clustered into  states. The main difference between this paper and our work is that our algorithm is able to model systems with non-linear behaviors.

\section{Conclusion and Future Work}
\label{sec:conclusion}

In this paper, we proposed an algorithm for learning a hybrid automaton from the input/output traces of a black-box system. We employed a similarity detection technique in an online manner to address the challenges related to clustering nonlinear behaviors of hybrid systems. The main feature of our proposed algorithm is its ability to perform in an online and passive way. Having online behavior allows the algorithm to work in real-time situations. With a passive technique, we don't face the limitations related to active interaction with the system under learning. Our experiments on a simulated hybrid system and a real-world safety-critical system show promising results.

As for the future work, we plan to identify systems with non-polynomial behaviors. Furthermore, in complex systems, jump conditions may be triggered by a complex predicate of different input signals. Our plan is to consider this type of systems for learning. The other limitation of this work is our assumption on synchronicity of the  system under learning. Mining  asynchronous behaviors is another line of future work.

\bibliography{ref}

\begin{thebibliography}{10}
\providecommand{\url}[1]{#1}
\csname url@samestyle\endcsname
\providecommand{\newblock}{\relax}
\providecommand{\bibinfo}[2]{#2}
\providecommand{\BIBentrySTDinterwordspacing}{\spaceskip=0pt\relax}
\providecommand{\BIBentryALTinterwordstretchfactor}{4}
\providecommand{\BIBentryALTinterwordspacing}{\spaceskip=\fontdimen2\font plus
\BIBentryALTinterwordstretchfactor\fontdimen3\font minus
  \fontdimen4\font\relax}
\providecommand{\BIBforeignlanguage}[2]{{%
\expandafter\ifx\csname l@#1\endcsname\relax
\typeout{** WARNING: IEEEtran.bst: No hyphenation pattern has been}%
\typeout{** loaded for the language `#1'. Using the pattern for}%
\typeout{** the default language instead.}%
\else
\language=\csname l@#1\endcsname
\fi
#2}}
\providecommand{\BIBdecl}{\relax}
\BIBdecl

\bibitem{medhat2015framework}
R.~Medhat, S.~Ramesh, B.~Bonakdarpour, and S.~Fischmeister, ``A framework for
  mining hybrid automata from input/output traces,'' in \emph{2015
  International Conference on Embedded Software (EMSOFT)}.\hskip 1em plus 0.5em
  minus 0.4em\relax IEEE, 2015, pp. 177--186.

\bibitem{soto2019membership}
M.~G. Soto, T.~A. Henzinger, C.~Schilling, and L.~Zeleznik, ``Membership-based
  synthesis of linear hybrid automata,'' in \emph{International Conference on
  Computer Aided Verification}.\hskip 1em plus 0.5em minus 0.4em\relax
  Springer, 2019, pp. 297--314.

\bibitem{jin2015mining}
X.~Jin, A.~Donz{\'e}, J.~V. Deshmukh, and S.~A. Seshia, ``Mining requirements
  from closed-loop control models,'' \emph{IEEE Transactions on Computer-Aided
  Design of Integrated Circuits and Systems}, vol.~34, no.~11, pp. 1704--1717,
  2015.

\bibitem{kong2014temporal}
Z.~Kong, A.~Jones, A.~Medina~Ayala, E.~Aydin~Gol, and C.~Belta, ``Temporal
  logic inference for classification and prediction from data,'' in
  \emph{Proceedings of the 17th international conference on Hybrid systems:
  computation and control}, 2014, pp. 273--282.

\bibitem{jha2019telex}
S.~Jha, A.~Tiwari, S.~A. Seshia, T.~Sahai, and N.~Shankar, ``Telex: learning
  signal temporal logic from positive examples using tightness metric,''
  \emph{Formal Methods in System Design}, vol.~54, no.~3, pp. 364--387, 2019.

\bibitem{raffelt2008hybrid}
H.~Raffelt, T.~Margaria, B.~Steffen, and M.~Merten, ``Hybrid test of web
  applications with webtest,'' in \emph{Proceedings of the 2008 workshop on
  Testing, analysis, and verification of web services and applications}, 2008,
  pp. 1--7.

\bibitem{aarts2015generating}
F.~Aarts, B.~Jonsson, J.~Uijen, and F.~Vaandrager, ``Generating models of
  infinite-state communication protocols using regular inference with
  abstraction,'' \emph{Formal Methods in System Design}, vol.~46, no.~1, pp.
  1--41, 2015.

\bibitem{fiteruau2016combining}
P.~Fiter{\u{a}}u-Bro{\c{s}}tean, R.~Janssen, and F.~Vaandrager, ``Combining
  model learning and model checking to analyze tcp implementations,'' in
  \emph{International Conference on Computer Aided Verification}.\hskip 1em
  plus 0.5em minus 0.4em\relax Springer, 2016, pp. 454--471.

\bibitem{weiss2018extracting}
G.~Weiss, Y.~Goldberg, and E.~Yahav, ``Extracting automata from recurrent
  neural networks using queries and counterexamples,'' in \emph{International
  Conference on Machine Learning}.\hskip 1em plus 0.5em minus 0.4em\relax PMLR,
  2018, pp. 5247--5256.

\bibitem{okudono2020weighted}
T.~Okudono, M.~Waga, T.~Sekiyama, and I.~Hasuo, ``Weighted automata extraction
  from recurrent neural networks via regression on state spaces,'' in
  \emph{Proceedings of the AAAI Conference on Artificial Intelligence},
  vol.~34, no.~04, 2020, pp. 5306--5314.

\bibitem{esparza2011learning}
J.~Esparza, M.~Leucker, and M.~Schlund, ``Learning workflow petri nets,''
  \emph{Fundamenta Informaticae}, vol. 113, no. 3-4, pp. 205--228, 2011.

\bibitem{choi2013automated}
W.~Choi, ``Automated testing of graphical user interfaces: A new algorithm and
  challenges,'' in \emph{Proceedings of the 2013 ACM workshop on Mobile
  development lifecycle}, 2013, pp. 27--28.

\bibitem{choi2013guided}
W.~Choi, G.~Necula, and K.~Sen, ``Guided gui testing of android apps with
  minimal restart and approximate learning,'' \emph{Acm Sigplan Notices},
  vol.~48, no.~10, pp. 623--640, 2013.

\bibitem{lamrani2018hymn}
I.~Lamrani, A.~Banerjee, and S.~K. Gupta, ``Hymn: mining linear hybrid automata
  from input output traces of cyber-physical systems,'' in \emph{2018 IEEE
  Industrial Cyber-Physical Systems (ICPS)}.\hskip 1em plus 0.5em minus
  0.4em\relax IEEE, 2018, pp. 264--269.

\bibitem{maier2014online}
A.~Maier, ``Online passive learning of timed automata for cyber-physical
  production systems,'' in \emph{2014 12th IEEE International Conference on
  Industrial Informatics (INDIN)}.\hskip 1em plus 0.5em minus 0.4em\relax IEEE,
  2014, pp. 60--66.

\bibitem{gold1978complexity}
E.~M. Gold, ``Complexity of automaton identification from given data,''
  \emph{Information and control}, vol.~37, no.~3, pp. 302--320, 1978.

\bibitem{pitt1993minimum}
L.~Pitt and M.~K. Warmuth, ``The minimum consistent dfa problem cannot be
  approximated within any polynomial,'' \emph{Journal of the ACM (JACM)},
  vol.~40, no.~1, pp. 95--142, 1993.

\bibitem{howar2018active}
F.~Howar and B.~Steffen, ``Active automata learning in practice,'' in
  \emph{Machine Learning for Dynamic Software Analysis: Potentials and
  Limits}.\hskip 1em plus 0.5em minus 0.4em\relax Springer, 2018, pp. 123--148.

\bibitem{henzinger2000theory}
T.~A. Henzinger, ``The theory of hybrid automata,'' in \emph{Verification of
  digital and hybrid systems}.\hskip 1em plus 0.5em minus 0.4em\relax Springer,
  2000, pp. 265--292.

\bibitem{bellman1959adaptive}
R.~Bellman and R.~Kalaba, ``On adaptive control processes,'' \emph{IRE
  Transactions on Automatic Control}, vol.~4, no.~2, pp. 1--9, 1959.

\bibitem{senin2008dynamic}
P.~Senin, ``Dynamic time warping algorithm review,'' \emph{Information and
  Computer Science Department University of Hawaii at Manoa Honolulu, USA},
  vol. 855, no. 1-23, p.~40, 2008.

\bibitem{truong2020selective}
C.~Truong, L.~Oudre, and N.~Vayatis, ``Selective review of offline change point
  detection methods,'' \emph{Signal Processing}, vol. 167, p. 107299, 2020.

\bibitem{kianimajd2017comparison}
A.~Kianimajd, M.~Ruano, P.~Carvalho, J.~Henriques, T.~Rocha, S.~Paredes, and
  A.~Ruano, ``Comparison of different methods of measuring similarity in
  physiologic time series,'' \emph{IFAC-PapersOnLine}, vol.~50, no.~1, pp.
  11\,005--11\,010, 2017.

\bibitem{le2018deep}
T.-D.~B. Le and D.~Lo, ``Deep specification mining,'' in \emph{Proceedings of
  the 27th ACM SIGSOFT International Symposium on Software Testing and
  Analysis}, 2018, pp. 106--117.

\bibitem{kang2021adversarial}
H.~J. Kang and D.~Lo, ``Adversarial specification mining,'' \emph{ACM
  Transactions on Software Engineering and Methodology (TOSEM)}, vol.~30,
  no.~2, pp. 1--40, 2021.

\bibitem{zhang2020temporal}
N.~Zhang, B.~Yu, C.~Tian, Z.~Duan, and X.~Yuan, ``Temporal logic specification
  mining of programs,'' \emph{Theoretical Computer Science}, 2020.

\bibitem{juloski2005comparison}
A.~L. Juloski, W.~Heemels, G.~Ferrari-Trecate, R.~Vidal, S.~Paoletti, and
  J.~Niessen, ``Comparison of four procedures for the identification of hybrid
  systems,'' in \emph{International Workshop on Hybrid Systems: Computation and
  Control}.\hskip 1em plus 0.5em minus 0.4em\relax Springer, 2005, pp.
  354--369.

\bibitem{lauer2019hybrid}
F.~Lauer and G.~Bloch, ``Hybrid system identification,'' in \emph{Hybrid System
  Identification}.\hskip 1em plus 0.5em minus 0.4em\relax Springer, 2019, pp.
  77--101.

\bibitem{blavzivc2020hybrid}
S.~Bla{\v{z}}i{\v{c}} and I.~{\v{S}}krjanc, ``Hybrid system identification by
  incremental fuzzy c-regression clustering,'' in \emph{2020 IEEE International
  Conference on Fuzzy Systems (FUZZ-IEEE)}.\hskip 1em plus 0.5em minus
  0.4em\relax IEEE, 2020, pp. 1--7.

\bibitem{pillonetto2016new}
G.~Pillonetto, ``A new kernel-based approach to hybrid system identification,''
  \emph{Automatica}, vol.~70, pp. 21--31, 2016.

\end{thebibliography}
\bibliographystyle{IEEEtran}

\end{document}